\documentclass[letterpaper, 10 pt, conference]{ieeeconf}  

\IEEEoverridecommandlockouts                              
\usepackage{amsmath,amssymb,amsthm,amsfonts,mathtools}
\usepackage{cite}
\usepackage{algorithmic}
\usepackage{graphicx}
\usepackage{bm}
\usepackage{multicol}
\usepackage{float}
\usepackage{xcolor}
\usepackage{subfigure}
\usepackage{textcomp}
\newtheorem{definition}{Definition}
\usepackage{url}
\let\labelindent\relax
\usepackage[inline]{enumitem}

\newtheorem{theorem}{Theorem}
\newtheorem{lemma}{Lemma}[theorem]
\newtheorem{corollary}{Corollary}[theorem]

\title{\LARGE \bf Uncertainty Propagation and Bayesian Fusion on Unimodular Lie Groups from a Parametric Perspective}
\author{Jikai Ye and Gregory S. Chirikjian
\thanks{This work was supported by National Research Foundation, Singapore, under its Medium Sized Centre Programme - Centre for Advanced Robotics Technology Innovation (CARTIN),  sub award A-0009428-08-00}
\thanks{J. Ye and G. S. Chirikjian are with the Department of Mechanical Engineering, National University of Singapore, Singapore (e-mail: \{jikai.ye, mpegre\}@nus.edu.sg).
G.S. Chirikjian is currently at the University of Delaware (e-mail: gchirik@udel.edu)}
}

\begin{document}

\maketitle
\thispagestyle{empty}
\pagestyle{empty}

\begin{abstract}
We address the problem of uncertainty propagation and Bayesian fusion on unimodular Lie groups.
Starting from stochastic differential equations (SDEs) defined on Lie groups via Mckean-Gangolli injection, we first convert to parametric SDEs in exponential coordinates.
The coefficient transform method for the conversion is stated for both Ito's and Stratonovich's interpretation of the SDE.
Then we derive a mean and covariance fitting formula for probability distributions on Lie groups defined by a concentrated distribution on the exponential coordinate.
It is used to derive the mean and covariance propagation equations for the SDE defined by injection, which coincides with the result derived from a Fokker-Planck equation in previous work.
We also propose a simple modification to the update step of Kalman filters using the fitting formula which improves the fusion accuracy.
\end{abstract}


\section{Introduction} \label{sec:intro}

In Bayesian filtering, the prediction step (uncertainty propagation) and update step (Bayesian fusion) execute iteratively to estimate the probability distribution of the state variable \cite{sarkka2023bayesian}.
When the state variable $\boldsymbol{x}$ lives in Euclidean space and white noise is present in the dynamic model, 
\begin{equation}
    d{\boldsymbol{x}}=\boldsymbol{h}(\boldsymbol{x},t)dt+H(t)d\boldsymbol{W},
\end{equation}
the evolution of the mean $\boldsymbol{\mu}(t)$ and covariance $\Sigma(t)$ of the state variable $\boldsymbol{x}$ is well-known \cite{sarkka2013gaussian} 
\begin{equation}\label{eq:euc_prop}
\begin{aligned}
   \begin{cases}
       \dot{\boldsymbol{\mu}}=\langle\boldsymbol{h} \rangle\\
     \dot{\Sigma}=\langle\boldsymbol{h}(\boldsymbol{x}-\boldsymbol{\mu})^T+(\boldsymbol{x}-\boldsymbol{\mu})\boldsymbol{h}^T\rangle+HH^T
    \end{cases}
\end{aligned}
\end{equation}
where $\langle\varphi \rangle \doteq \mathbb{E}(\varphi)$.
However, when the state space has a Lie group structure, the mean and covariance propagation process is much less understood.
This paper derives a propagation equation in this setting from a stochastic differential equation perspective, which complements and provides more intuition to the propagation equation derived from Fokker-Planck equations on unimodular Lie groups in \cite{ye2023uncertainty}.
One intermediate result in the derivation motivates a simple modification term to the update step of the Gaussian filter that improves the Bayesian fusion accuracy.

When the state variable lives in a Lie group, the noisy dynamic model is a stochastic differential equation (SDE) on the Lie group.
There exist several ways to define such an SDE \cite{chirikjian2011stochastic}.
In \cite{solo2019ito,marjanovic2018numerical,li2023errors}, a matrix Lie group is considered as a submanifold embedded in $GL(n)$, and the SDE is defined on $GL(n)$ directly. 
Another way is to use the McKean-Gangolli injection \cite{mckean1969stochastic,chirikjian2011stochastic,brossard2017unscented} in which an infinitesimal stochastic process on the Lie algebra is projected to the Lie group via the exponential map.
These two types of SDEs can be interconverted if the coefficients satisfy a simple relationship \cite{marjanovic2016engineer,solo2019ito}.
A third way is to parametrize the group and define an SDE on the parameter space.
In this paper, we start with the second type of SDE and convert it to the third type.
When using exponential coordinates as the parametrization and Stratonovich's interpretation, the coefficient conversion is very simple, while when using Ito's interpretation, an additional drift term appears.
Note that the Fokker-Planck equation corresponding to these two equivalent SDEs is derived in \cite{ye2023uncertainty}, which describes the evolution of the probability density function of the state variable.

We proceed to review the mean and covariance propagation methods.
In extended Kalman-filter-like methods, the mean is propagated using the dynamic model without noise \cite{barrau2016invariant,barrau2018invariant,ge2023note,bourmaud2015continuous,bourmaud2016intrinsic} and the propagation of covariance is derived by expansion and truncation.
In the literature that employs unscented transform, the mean and covariance propagation can be calculated by (i) propagating sigma points on the group and performing optimization \cite{forbes2017sigma,hauberg2013unscented}; (ii) propagating sigma points on the tangent space and projecting the mean and covariance back to the group \cite{loianno2016visual,hauberg2013unscented,sjoberg2021lie}; or (iii) propagating the mean using the deterministic dynamic model \cite{brossard2017unscented,brossard2020code} and the covariance by unscented transform.
In this work, we take the second approach which utilizes the tangent space for propagation.
Different from previous work that projects mean and covariance alone, we project quantities on the tangent space back to the group while taking into account the influence of the probability distribution.
Also, we minimize the approximations in the derivation and arrive at a propagation equation in the form of (\ref{eq:euc_prop}) assuming the initial probability distribution is concentrated.
When the initial probability distribution is non-zero but the noise is zero, our propagation equation is exact.
A propagation equation of the same form has been derived previously in \cite{ye2023uncertainty} from a Fokker-Planck equation's perspective and the approximate equation based on it has been demonstrated by experiments.
We offer another derivation that provides more information about the meaning of each term in the equation from an SDE perspective.
The difference between the propagation equation in this paper and the one in \cite{ye2023uncertainty} comes from the choice of right/left SDEs.

To fuse a new observation with a prior distribution on Lie groups, various approaches exist.
Kalman filter-based methods \cite{bourmaud2015continuous,barrau2016invariant,barrau2018invariant,brossard2017unscented,hauberg2013unscented} estimate the mean and covariance of the posterior distribution on the tangent space and project them back to the group.
In \cite{barfoot2017state,bourmaud2016intrinsic}, the fusion problem is considered as finding the mode of the posterior distribution and is solved by optimization algorithms.
Using explicit polynomial expansion and coefficient matching, the fusion problem can be transformed into a set of algebraic equations that has approximate solutions of different order as in \cite{wolfe2011bayesian,chirikjian2014gaussian}.
When the observation lives in a manifold, the authors in \cite{ge2023note} raise the issue of parallel transport of covariance due to the curved nature of the manifold.
For applications that consider large covariance, Fourier analysis on Lie groups can also be used \cite{kim2015bayesian}.
In our work, we adopt the Kalman filter update approach due to its simplicity and fast computation.
A modification term motivated by the derivation of the propagation equation is added to the projection step from the tangent space to the group, which improves the fusion's accuracy.

\textit{Contributions}: 
i) We state the equivalence between a non-parametric stochastic differential equation (SDE) on a Lie group defined by Mckean-Gangolli injection and a parametric SDE defined on exponential coordinates of the Lie group.
ii) A formula with error analysis is derived for fitting the group-theoretic mean and covariance of a probability defined on exponential coordinates of a Lie group.
iii) A continuous-time mean and covariance propagation equation and a Bayesian fusion method are derived using exponential coordinates and the mean and covariance fitting method.


\section{Background} \label{sec:back}

In this section, we provide a minimal introduction to Lie group theory.
For a comprehensive introduction, please refer to \cite{chirikjian2011stochastic}.
The \textit{Einstein summation} convention is used to simplify notations throughout this paper.

An $N$-dimensional matrix Lie group $G$ is an $N$-dimensional analytic manifold and also a subgroup of the general linear matrix $GL(n)$ with group product and inverse operation being analytic.
The Lie algebra $\mathcal{G}$ of a Lie group is the tangent space at the identity of $G$ equipped with a Lie bracket.
In the case of a $N$-dimensional matrix Lie group, its Lie algebra $\mathcal{G}$ can be understood as a $N$-dimensional linear space consisting of matrices whose matrix exponential are in $G$ and the Lie bracket is defined by
\begin{equation}
    [X,Y]\doteq XY-YX, \,\,X, Y\in \mathcal{G}.
\end{equation}
Given a basis of $\mathcal{G}$ as $\{E_i\}_{i=1,2,...,N}$, we can draw equivalence between the Lie algebra $\mathcal{G}$ and $\mathbb{R}^N$ using the `$\wedge$' and `$\vee$' operation: $\boldsymbol{x}^{\wedge}\doteq \sum_{i=1}^N x_iE_i\in \mathcal{G}, \,\, \boldsymbol{x}\in \mathbb{R}^N$ and $X^{\vee}=\boldsymbol{x}\in \mathbb{R}^N$ which is the inverse of `$\wedge$'.
This identification of $E_i$ with $\boldsymbol{e}_i$ is equivalent to fixing a metric for $G$.
The little `$ad$' operator is defined by $ad_{X}Y\doteq[X,Y],\,\,X,Y\in \mathcal{G}$.
This operator is a linear operator on $Y$ and can be transformed into a matrix $[ad_X]\in \mathbb{R}^{N\times N}$ that satisfies $[ad_X]\boldsymbol{y}=(ad_{X}Y)^{\vee}$.

Since a matrix Lie group is also a manifold, we can locally parametrize it by a subset of $\mathbb{R}^N$ as $g(\boldsymbol{q})\in G$ where $\boldsymbol{q}\in \mathbb{R}^N$.
One parametrization that exists for all matrix Lie groups is the \textit{exponential coordinate}, where parametrization around $\mu \in G$ is obtained by the matrix multiplication and matrix exponential, $g(\boldsymbol{x})=\mu \exp(\boldsymbol{x}^{\wedge})$.
The neighborhood of any group element $\mu\in G$ can be parametrized in this way and the parametrization is a local diffeomorphism between $D\subseteq \mathbb{R}^N$ and $G$.
The domain of exponential coordinates, $D$, is specified case by case, for example $D=\{\boldsymbol{x}\in \mathbb{R}^3 \,|\, \lVert\boldsymbol{x}\rVert_2 <\pi \}$ for $SO(3)$.
The expansion of the Jacobian matrix (and its inverse) corresponding to this parametrization is known to be a power series of `$ad$' \cite{bullo1995proportional}.

The left and right Jacobian matrices of $G$ are defined by
\begin{equation}\small \label{eq:l_jac}
    J_l(\boldsymbol{q})=\bigg[\bigg(\frac{\partial g(\boldsymbol{q})}{\partial q_1}g^{-1}\bigg)^{\vee},\bigg(\frac{\partial g(\boldsymbol{q})}{\partial q_2}g^{-1}\bigg)^{\vee},...,\bigg(\frac{\partial g(\boldsymbol{q})}{\partial q_N}g^{-1}\bigg)^{\vee}\bigg],
\end{equation}
\begin{equation}\small \label{eq:r_jac}
    J_r(\boldsymbol{q})=\bigg[\bigg(g^{-1}\frac{\partial g(\boldsymbol{q})}{\partial q_1}\bigg)^{\vee},\bigg(g^{-1}\frac{\partial g(\boldsymbol{q})}{\partial q_2}\bigg)^{\vee},...,\bigg(g^{-1}\frac{\partial g(\boldsymbol{q})}{\partial q_N}\bigg)^{\vee}\bigg],
\end{equation}
where $g(\boldsymbol{q})$ can be the exponential or any other parametrization.
They are useful in computing the left and right Lie directional derivatives defined by 
\begin{equation*} \label{eq:l_lie}
    E_i^l f(g)\doteq \frac{d}{dt}f(\exp(-tE_i)g)=-[J_l^{-T}(\boldsymbol{q})]_{ij}\frac{\partial f(g(\boldsymbol{q}))}{\partial q_j},
\end{equation*}
\begin{equation*} \label{eq:r_lie}
    E_i^r f(g)\doteq \frac{d}{dt}f(g\exp(tE_i))=[J_r^{-T}(\boldsymbol{q})]_{ij}\frac{\partial f(g(\boldsymbol{q}))}{\partial q_j},
\end{equation*}
which are generalizations of partial derivatives on Euclidean space. 
The `$l$' and `$r$' denote on which side a perturbation is applied to the argument $g$.
Taylor's expansion on Euclidean space can be generalized to Lie groups using Lie directional derivative
\begin{equation*}
    f(g\exp(\boldsymbol{\epsilon}^{\wedge}))=f(g)+\epsilon_i E^r_i(f)+\frac{1}{2}\epsilon_i \epsilon_j E_i^r E_j^r(f)+\mathcal{O}(\lVert \boldsymbol{\epsilon} \rVert^3)
.
\end{equation*}
Another use of Jacobians is to solve ordinary differential equations on $G$ by parametrization.
For example,
\begin{equation}\label{eq:deter_ode}
    (g^{-1}\dot{g})^{\vee}=\boldsymbol{h}(g) \iff \dot{\boldsymbol{q}}=J_r^{-1}(\boldsymbol{q})\boldsymbol{h}(g(\boldsymbol{q}))
\end{equation}
when the Jacobian matrix is not singular.
The Jacobian can also be used to construct two Haar measures on $G$ defined by $d_l g\doteq |\det J_l|d\boldsymbol{q}$ and $d_r g\doteq |\det J_r|d\boldsymbol{q}$.
When $|\det J_l|\equiv |\det J_r|$, the group is called unimodular and the two Haar measures are both invariant to left and right shift.

The integration of a function on an unimodular Lie group $G$ should be formally defined using the Haar measure and partition of unity.
In this paper, we only consider functions whose support is within the domain of exponential coordinates centered at some group element $\mu$, i.e. $\text{supp}(f)\subseteq \mu\exp(D)$, and the integration can be calculated on the exponential coordinate by
\begin{equation}
    \int_G f(g)dg\doteq \int_D f(\mu \exp(\boldsymbol{x}))|J_r(\boldsymbol{x})|d\boldsymbol{x}.
\end{equation}
To simplify equations, we notationally suppress the `$\wedge$' operator in the exponential map, which we will continue to use throughout this paper.
For estimation problems, the probability density function of the state variable generally falls within this category.

A probability density function $p(g)$ on $G$ is a function that satisfies: i) $p(g)\geq0$ and ii) $\int_G p(g)dg=1$. 
Building on terminology already in use in literature \cite{chirikjian2011stochastic,sjoberg2021lie,bourmaud2015continuous}, we now formally define the concept of \textit{concentrated distribution} on $\mathbb{R}^N$:
\begin{definition}
    If a probability distribution on $\mathbb{R}^N$ satisfies the following properties, it is said to be a concentrated distribution near the origin of domain $D$: i) the support of the probability density function $\tilde{p}(\boldsymbol{x})$ is within domain $D$, i.e. $\text{supp}(\tilde{p})\subseteq D$, ii) the mean of the distribution is around the origin, i.e. $\lVert\mathbb{E}(\boldsymbol{x})\rVert_2 \ll 1$, and iii) the covariance matrix $\Sigma$ of the distribution is small, i.e. $\lVert \Sigma \rVert_2 \ll 1$.
\end{definition}
\noindent Given a concentrated distribution near the origin of $D$, $\tilde{p}(\boldsymbol{x})$, we can construct a probability distribution on any unimodular Lie group by defining a random variable $g(\boldsymbol{x})=\mu \exp(\boldsymbol{x})\in G$.
Denote the probability density function of this random variable as $p(g)$.
The expectation of a function $f(g)$ can be calculated as
\begin{equation}
    \langle f\rangle\doteq\int_G f(g)p(g)dg=\int_{\mathbb{R}^N} f(\mu\exp(\boldsymbol{x}))\tilde{p}(\boldsymbol{x})d\boldsymbol{x}
\end{equation}
where 
\begin{equation}
p(\mu\exp(\boldsymbol{x}))|J_r(\boldsymbol{x})|=\tilde{p}(\boldsymbol{x}).
\end{equation}
One core contribution of this paper is to estimate the mean and covariance of $p(g)$ defined below.

Given a probability density function $p(g)$ on an unimodular Lie group, we may define two types of mean.
One is the \textit{group-theoretic mean} defined by \cite{chirikjian2011stochastic}
\begin{equation}
     \int_G \log^{\vee}(\mu_{G}^{-1}g)p(g)dg = \boldsymbol{0}
\end{equation}
 and another is the \textit{Fr\'echet mean} defined by \cite{hauberg2013unscented}
\begin{equation}
    \mu_{F} = \text{argmin}_{\mu} \int_G d^2(\mu,g)p(g)dg.
\end{equation}
These two definitions are not equivalent in general, but in some situations such as the one stated below, they are the same:
\begin{theorem} \label{thm:mean_eq}
    When the group is $SO(3)$ and the distance is chosen as $d(\mu,g)=|\log^{\vee}(\mu^{-1}g)|_2$, the \textit{Fr\'echet mean} and the \textit{group-theoretic} mean are the same.
    \begin{proof}
        It is proved using the first-order condition for the minimization problem and the structure of the Jacobian of $SO(3)$.
    \end{proof}
\end{theorem}
\noindent The covariance of $p(g)$ is defined by
\begin{equation}
    \Sigma=\int_G [\log^{\vee}(\mu^{-1}g)][\log^{\vee}(\mu^{-1}g)]^T p(g)dg
\end{equation}
where $\mu$ is the mean of $p(g)$.
In this paper, we only consider the \textit{group-theoretic} mean and covariance.


\section{Stochastic Differential Equations on Lie Groups}
Suppose we have a N-dimensional matrix Lie group, $G$, a vector-valued function, $\boldsymbol{h}:G\times \mathbb{R}\rightarrow \mathbb{R}^N$, and a matrix-valued function, $H:G\times \mathbb{R}\rightarrow \mathbb{R}^{N\times N}$. Denote an $N$-dimensional Wiener process as $\boldsymbol{W}(t)$ which satisfies $\big(\boldsymbol{W}(t+s)-\boldsymbol{W}(t)\big)\sim \mathcal{N}(\boldsymbol{0},s\mathbb{I}_{N\times N})$. 
In the paper, we notationally suppress the `$\wedge$' operator in the exponential map and directly write $\exp(\boldsymbol{x})$ for simplicity.

\begin{definition} (non-parametric SDE on $G$) A stochastic differential equation on $G$ can be defined non-parametrically via Mckean-Gangolli injection \cite{mckean1969stochastic},
\begin{equation}
g(t+dt)=g(t)\exp\big(\boldsymbol{h}\Bigr|_{\tiny\substack{g=g(t)\\t=t}}dt+H\Bigr|_{\tiny\substack{g=g(t+\kappa dt)\\t=t+\kappa dt}}d\boldsymbol{W}\big),\label{eq:group_SDE}
\end{equation}
where $dt$ is the infinitesimal increment of time and $d\boldsymbol{W}\doteq \boldsymbol{W}(t+dt)-\boldsymbol{W}(t)$.
\end{definition}
\noindent The SDE is called Ito's when $\kappa=0$ and Stratonovich's when $\kappa=\frac{1}{2}$, which is consistent with the definitions on Euclidean space \cite{gardiner1985handbook}.  
We also use the following simplified notation to denote the same SDE
$$
g^{-1}dg=\boldsymbol{h}dt+Hd\boldsymbol{W}
$$
where we use $Hd\boldsymbol{W}$ to denote a Ito's SDE and $H\circledS d\boldsymbol{W}$ for a Stratonovich's SDE as in \cite{ye2023uncertainty}. 
A sample path of the non-parametric SDE (\ref{eq:group_SDE}) starting at $g(0)$ is defined by the following limit
\begin{equation}\label{eq:sol_group_SDE}\small
\begin{aligned}
g(T)\!\!&=\!\!\lim_{M\rightarrow \infty}\!\! g(0)\!\!\prod_{i=0}^{M-1} \!\! \exp\bigg[\boldsymbol{h}\Bigr|_{\tiny\substack{g=g(t_i)\\t=t_i}} \Delta t+H\Bigr|_{\tiny\substack{g=g(t_i+\kappa \Delta t)\\t=t_i+\kappa \Delta t}}\Delta \boldsymbol{W}_i\bigg]
\end{aligned}
\end{equation}
where $\Delta t=T/M$, $t_i=i \Delta t$, $\Delta \boldsymbol{W}_i=\boldsymbol{W}(t_{i+1})-\boldsymbol{W}(t_{i})$, and the exponential of increment is multiplied on the right sequentially.

Another way to define an SDE on Lie groups is to parameterize group elements and define an SDE on the parameter space:

\begin{definition}\label{def:para_SDE}
(parametric SDE on $G$) A stochastic differential equation on $G$ can be defined parametrically by parametrizing group elements as $g=g(\boldsymbol{q})$ and writing an SDE on the parameter space as
\begin{equation}\small
\begin{aligned}
    \boldsymbol{q}(t+dt)\!&=\!\boldsymbol{q}(t)+(J_r^{-1}\tilde{\boldsymbol{h}})\Bigr|_{\tiny\substack{\boldsymbol{q}=\boldsymbol{q}(t)\\t=t}}dt\!+\!(J_r^{-1}\tilde{H})\Bigr|_{\tiny\substack{\boldsymbol{q}=\boldsymbol{q}(t+\kappa dt)\\t=t+\kappa dt}}d\boldsymbol{W}
    \end{aligned}\label{eq:para_SDE}
\end{equation}
where $\tilde{\boldsymbol{h}}(\boldsymbol{q},t)$ and $\tilde{H}(\boldsymbol{q},t)$ are functions of the parameters and time. 
\end{definition} 
\noindent As before, it is called Ito's when $\kappa=0$ and Stratonovich's when $\kappa =1/2$. 
For a short time, assume the trajectory $\boldsymbol{q}(t)$ is still in the domain of the parametrization, a sample path of the SDE starting at $\boldsymbol{q}(0)$ is defined by the following limit \cite{gardiner1985handbook}
\begin{equation}\small
    \begin{aligned}
\boldsymbol{q}(&T)=\boldsymbol{q}(0)+\\&\lim_{M\rightarrow \infty}\!\sum_{i=0}^{M-1}\!\bigg\{(J_r^{-1}\tilde{\boldsymbol{h}})\Bigr|_{\tiny\substack{\boldsymbol{q}=\boldsymbol{q}(t_i)\\t=t_i}} \Delta t\!+\!(J_r^{-1}\tilde{H})\Bigr|_{\tiny\substack{\boldsymbol{q}=\boldsymbol{q}(t_i+\kappa \Delta t)\\t=t_i+\kappa \Delta t}}\Delta\boldsymbol{W}_i\bigg\}
\end{aligned}\label{eq:sol_para_SDE}
\end{equation}
where $\Delta t=T/M$, $t_i=i \Delta t$ and $\Delta \boldsymbol{W}_i=\boldsymbol{W}(t_{i+1})-\boldsymbol{W}(t_i)$.
The path is then mapped back to $G$ by $g(t)=g(\boldsymbol{q}(t))$.

When using exponential coordinates to parametrize the group, i.e. $g(\boldsymbol{x})=\mu \exp(\boldsymbol{x})$, the sample paths of equation (\ref{eq:group_SDE}) and equation (\ref{eq:para_SDE}) are related by the following theorem:

\begin{theorem}\label{thm:Ito_equi} Using the parametrization $g(\boldsymbol{x})=\mu \exp(\boldsymbol{x})$ in Definition \ref{def:para_SDE}, when both equation (\ref{eq:group_SDE}) and equation (\ref{eq:para_SDE}) are interpreted as Ito's SDEs, i.e. $\kappa=0$, their sample paths are equivalent if the following condition holds
\begin{equation}
\begin{aligned}
\tilde{\boldsymbol{h}}\Bigr|_{\tiny\substack{\boldsymbol{x}=\boldsymbol{x}\\t=t}}&=\boldsymbol{h}\Bigr|_{\tiny\substack{g=\mu \exp(\boldsymbol{x})\\t=t}}\!\!+\!\bigg(\frac{1}{2}J_r\frac{\partial J_r^{-1}}{\partial x_k} H  H^T J_r^{-T}\bigg)\Biggr|_{\tiny\substack{g=\mu \exp(\boldsymbol{x})\\
\boldsymbol{x}=\boldsymbol{x}\\t=t}}\boldsymbol{e}_k,\\
&\qquad \qquad\tilde{H}\Bigr|_{\tiny\substack{\boldsymbol{x}=\boldsymbol{x}\\t=t}}=H\Bigr|_{\tiny\substack{g=\mu\exp(\boldsymbol{x})\\t=t}}.
\end{aligned}
\end{equation}
\begin{proof}
At time $t$ and $t+dt$, we parametrize group elements by $g=\mu\exp(\tilde{\boldsymbol{x}}(t))$ and $g=\mu\exp(\tilde{\boldsymbol{x}}(t+d t))$.
Using the definition of the non-parametric SDE, we have
\begin{equation}\small
    \begin{aligned}
&\,\,\quad\mu\exp(\tilde{\boldsymbol{x}}(t+dt))\\
&=\mu\exp(\tilde{\boldsymbol{x}}(t))\exp\big[\boldsymbol{h}\Bigr|_{\tiny\substack{g=\mu\exp(\tilde{\boldsymbol{x}}(t))\\t=t}}dt+H\Bigr|_{\tiny\substack{g=\mu\exp(\tilde{\boldsymbol{x}}(t))\\t=t}}d\boldsymbol{W}\big]\\
&=\mu\exp\!\big[\tilde{\boldsymbol{x}}(t)\!+\!\big(J_r^{-1}\boldsymbol{h}\Bigr|_{\tiny\substack{g=\mu\exp(\tilde{\boldsymbol{x}}(t))\\t=t}}dt+J_r^{-1}H\Bigr|_{\tiny\substack{g=\mu \exp(\tilde{\boldsymbol{x}}(t))\\t=t}}\!d\boldsymbol{W}\big)\\
&\qquad+\frac{1}{2}\frac{\partial J_r^{-1}}{\partial x_k}\big(\boldsymbol{h}\Bigr|_{\tiny\substack{g=\mu\exp(\tilde{\boldsymbol{x}}(t))\\t=t}}dt+H\Bigr|_{\tiny\substack{g=\mu\exp(\tilde{\boldsymbol{x}}(t))\\t=t}}d\boldsymbol{W}\big) \\
&\qquad\quad\cdot\big(\boldsymbol{h}\Bigr|_{\tiny\substack{g=\mu\exp(\tilde{\boldsymbol{x}}(t))\\t=t}}dt \quad\!\!+\!H\Bigr|_{\tiny\substack{g=\mu\exp(\tilde{\boldsymbol{x}}(t))\\t=t}}d\boldsymbol{W}\big)^TJ_r^{-T}\!\boldsymbol{e}_k\\
&\qquad\qquad+O(dt^{3/2})\big]\\
&=\mu\exp\bigg[\tilde{\boldsymbol{x}}(t)\!+\!\bigg(\!J_r^{-1}\boldsymbol{h}\!+\!\frac{1}{2}\frac{\partial J_r^{-1}}{\partial x_k} H  H^T J_r^{-T}\boldsymbol{e}_k\bigg)\Biggr|_{\tiny\substack{g=\mu\exp(\tilde{\boldsymbol{x}}(t))\\ \boldsymbol{x}=\tilde{\boldsymbol{x}}(t) \\t=t}}dt\\
&\qquad\qquad+(J_r^{-1}H)\Biggr|_{\tiny\substack{g=\mu\exp(\tilde{\boldsymbol{x}}(t))\\ \boldsymbol{x}=\tilde{\boldsymbol{x}}(t)\\t=t}}d\boldsymbol{W}+O(dt^{3/2})\bigg]
\end{aligned}
\end{equation}
where we have used $dW_i \cdot dW_j =\delta_{ij}dt$, $dW_i dt \sim \mathcal{O}(dt^{3/2})$ and the expansion formula in the Appendix.
Comparing it with the parametric SDE, we arrive at the condition.
\end{proof}

\end{theorem}
\noindent \textit{Remark}: At first glance, it is surprising that an additional appears when using a parametric SDE to describe a non-parametric SDE.
That term comes from the non-linearity of the exponential map which is used in defining the non-parametric SDE. Previous work \cite{marjanovic2018numerical,li2023errors} also observe a similar phenomenon in their definitions of SDE.

\begin{lemma} \label{lemma:Ito_Stra_equi} The following non-parametric Ito's and Stratonovich's SDEs,
\begin{equation}
\begin{aligned}
(g^{-1}dg)^{\vee}&=\boldsymbol{h}dt+Hd\boldsymbol{W},\\
(g^{-1}dg)^{\vee}&=\boldsymbol{h}^sdt+H^s \circledS d\boldsymbol{W},
\end{aligned}
\end{equation}
are equivalent when
\begin{equation}
\boldsymbol{h}=\boldsymbol{h}^s+\frac{1}{2}E^r_i({H}^s_{kj}){H}^s_{ij}\boldsymbol{e}_k.
\end{equation}
\begin{proof}
    The proof is in the Appendix. 
\end{proof}
\end{lemma}

\begin{theorem} Using the parametrization $g=\mu \exp(\boldsymbol{x})$ in equation (\ref{eq:para_SDE}), when both the non-parametric and the parametric SDEs are interpreted as Stratonovich's SDEs, i.e. $\kappa=1/2$, their solutions are equivalent when
\begin{equation}
\begin{aligned}
\tilde{\boldsymbol{h}}^s\Bigr|_{\tiny\substack{\boldsymbol{x}=\boldsymbol{x}\\t=t}}&=\boldsymbol{h}^s\Bigr|_{\tiny\substack{g=\mu\exp(\boldsymbol{x})\\t=t}}\quad\text{and}\quad \tilde{H}^s\Bigr|_{\tiny\substack{\boldsymbol{x}=\boldsymbol{x}\\t=t}}=H^s\Bigr|_{\tiny\substack{g=\mu\exp(\boldsymbol{x})\\t=t}}.
\end{aligned}
\end{equation}
\begin{proof}
    Combining Theorem \ref{thm:Ito_equi} and Lemma \ref{lemma:Ito_Stra_equi}, we will have the result.
\end{proof}
\end{theorem}

\noindent \textit{Remark}: The connection between a parametric and a non-parametric SDE on $G$ is natural when both are interpreted as Stratonovich's.  \label{sec:sde_lie_groups}


\section{Propagation and Fusion on Lie Groups} \label{sec:gaussian_filtering}

In this section, we provide a derivation for the mean and covariance propagation equations on unimodular Lie groups from an SDE perspective and a modification to the update step of a traditional Gaussian/Kalman filter.
We first present a theorem that estimates the group-theoretic mean and covariance of a probability distribution on the exponential coordinate with error analysis in Section A.
It is then used to derive a continuous-time mean and covariance propagation equation in Section B.
A Bayesian fusion method using the mean estimation theorem is presented in Section C.

\subsection{Mean and Covariance Fitting for Concentrated Distribution on Exponential Coorinates}
The following theorem gives an estimation of the group-theoretic mean and covariance of the random variable $g(\boldsymbol{x})=\mu \exp(\boldsymbol{x})$ when $\boldsymbol{x}$ obeys a concentrated distribution.
The key observation is that even if the mean $\boldsymbol{m}=\mathbb{E}(\boldsymbol{x})$ is infinitesimal, the resulting group-theoretic mean $\mu_m$ is not exactly $\mu \exp(\boldsymbol{m})$.
\begin{theorem}\label{thm:fit}
    Given a random variable $\boldsymbol{x}\in\mathbb{R}^N$ whose probability distribution is concentrated around the origin of $D$.
    Denote its mean, covariance matrix, and the probability density function by $\boldsymbol{m}$, $\Sigma$, and $\tilde{p}(\boldsymbol{x})$.
    The random variable defined by $g=\mu\exp(\boldsymbol{x})$ obeys a distribution whose group-theoretic mean $\mu_m$ and covariance $\Sigma_m$ are estimated by
    \begin{equation}
        \begin{aligned}
        &\boldsymbol{m}'=\langle J_l^{-1} \rangle^{-1} \boldsymbol{m} \\
        &\mu_m = \mu\exp (\boldsymbol{m}'+\mathcal{O}(|\boldsymbol{m}'|^2))\\
        &\Sigma_{m}=\Sigma-\text{sym}\big(\langle J_l^{-1}\boldsymbol{m}'\boldsymbol{x}^T \rangle\big)
+\mathcal{O}(|\boldsymbol{m}'|^2)
        \end{aligned}
    \end{equation}
    where $\langle v(\boldsymbol{x})\rangle\doteq \int_{\mathbb{R}^N} v(\boldsymbol{x})\tilde{p}(\boldsymbol{x})d\boldsymbol{x}$ and $sym(A)\doteq A+A^T$.
    \begin{proof}
    The group-theoretic mean of $g$ can be written in the form of $\mu_m = \mu \exp(\boldsymbol{\epsilon})$.
    By definition, the mean should satisfy
    \begin{equation}
        \boldsymbol{0}=\int_{\mathbb{R}^N} \log^{\vee}\big( \exp(-\boldsymbol{\epsilon})\exp(\boldsymbol{x}) \big)\tilde{p}(\boldsymbol{x})d\boldsymbol{x}.
    \end{equation}  
    Using the expansion formula in the Appendix,
    \begin{equation}
        \begin{aligned}
\log^{\vee}(\exp(\boldsymbol{-\epsilon}^{\wedge})\exp(\boldsymbol{x}^{\wedge}))=\boldsymbol{x}-J_l^{-1}(\boldsymbol{x})\boldsymbol{\epsilon}+\mathcal{O}(\boldsymbol{\epsilon}^2),
\end{aligned}
    \end{equation}
    we have
    \begin{equation}
    \begin{aligned}
         \boldsymbol{\epsilon}&=\langle J_l^{-1}\rangle^{-1} \langle \boldsymbol{x} \rangle + \mathcal{O}(|\boldsymbol{\epsilon}|^2)\\
         &=\langle J_l^{-1}\rangle^{-1}  \boldsymbol{m}  + \mathcal{O}(|\boldsymbol{\epsilon}|^2),
         \end{aligned}
    \end{equation}
    which indicates $\boldsymbol{\epsilon}$ is of the same order as $\boldsymbol{m}$.
    So we can write
        \begin{equation}
    \begin{aligned}
         \boldsymbol{\epsilon}&= \boldsymbol{m}' + \mathcal{O}(|\boldsymbol{m}'|^2) \quad \text{and} \quad \boldsymbol{m}'=\langle J_l^{-1}\rangle^{-1} \boldsymbol{m}.
         \end{aligned}
    \end{equation}
    For the covariance matrix, we have
    \begin{equation}
        \begin{aligned}
            \begin{aligned}
\Sigma_m&=\big\langle\log^{\vee}[g_{\epsilon}^{-1}\exp(\boldsymbol{x})]{\log^{\vee}}^T[g_{\epsilon}^{-1}\exp(\boldsymbol{x})]\big\rangle\\
&=\big\langle \big[\boldsymbol{x}-J_l^{-1}\boldsymbol{\epsilon}+\mathcal{O}(|\boldsymbol{\epsilon}|^2)\big] \big[\boldsymbol{x}-J_l^{-1}\boldsymbol{\epsilon}+\mathcal{O}(|\boldsymbol{\epsilon}|^2)\big]^T \big\rangle\\
&=\Sigma-\big\langle\text{sym}[J_l^{-1}\boldsymbol{\epsilon}\boldsymbol{x}^T]\big \rangle+\mathcal{O}(|\boldsymbol{\epsilon}|^2)\\
&=\Sigma-\big\langle\text{sym}[J_l^{-1}\boldsymbol{m}'\boldsymbol{x}^T]\big \rangle+\mathcal{O}(|\boldsymbol{m}'|^2).
\end{aligned}
        \end{aligned}
    \end{equation}
    \end{proof}
\end{theorem}

Next, we apply this theorem to derive a continuous-time mean and covariance propagation equation for a non-parametric SDE on Lie groups.

\subsection{Propagation Equations}\label{sec:prop_eqs}
Suppose we have a stochastic process on Lie group $G$ described by the following non-parametric Ito's SDE,
\begin{equation}\label{eq:prop_nonpara_SDE}
g(t+dt)=g(t)\exp\big(\boldsymbol{h}\Bigr|_{\tiny\substack{g=g(t)\\t=t}}dt+Hd\boldsymbol{W}\big),
\end{equation}
where $\boldsymbol{h}:G\times \mathbb{R}\rightarrow \mathbb{R}^N$ and $H$ is a $N\times N$ constant matrix. 
We aim to derive the propagation equations for the group-theoretic mean $\mu(t)$ and covariance $\Sigma(t)$ of $g(t)$.
At time $t$, we parametrize group elements by $g(t)=\mu(t)\exp(\boldsymbol{x}(t))$ and write the probability density function of $\boldsymbol{x}$ as $\tilde{p}(\boldsymbol{x},t)$. 
We assume $\tilde{p}(\boldsymbol{x},t)$ to be a concentrated distribution around the origin of $D$.
To simplify equations, we use the notation
\begin{equation}
\langle v(\boldsymbol{x})\rangle \doteq \int_{\mathbb{R}^N}v(\boldsymbol{x})\tilde{p}(\boldsymbol{x},t)d\boldsymbol{x}.    
\end{equation}

\begin{theorem}
    The group-theoretic mean $\mu(t)$ and covariance $\Sigma(t)$ of a stochastic process $g(t)$ described by the SDE (\ref{eq:prop_nonpara_SDE}) obey the following ordinary differential equations:
    \begin{equation}
        (\mu^{-1}\dot{\mu})^{\vee}{\approx}\langle J_l^{-1}\rangle^{-1}\bigg\langle \frac{1}{2}\frac{\partial J_r^{-1}}{\partial x_k}(HH^T J_r^{-T}\boldsymbol{e}_k)+J_r^{-1}\boldsymbol{h}^c \bigg\rangle
    \end{equation}
    and
    \begin{equation}
        \begin{aligned}
\dot{\Sigma}&\,{\approx}\,\bigg\langle \text{sym}\bigg[(\frac{1}{2}\frac{\partial J_r^{-1}}{\partial x_k}(HH^TJ_r^{-T})\boldsymbol{e}_k-J_l^{-1}(\mu^{-1}\dot{\mu})^{\vee}\\&\qquad\qquad+J_r^{-1}\boldsymbol{h}^c)\boldsymbol{x}^T \bigg]+J_r^{-1}HH^T J_r^{-T}\bigg\rangle.
\end{aligned}
    \end{equation}
    where $\boldsymbol{h}^c(\boldsymbol{x},t)\doteq\boldsymbol{h}(\mu\exp(\boldsymbol{x}),t)$ and $sym(A)\doteq A+A^T$.
    In the case where $HH^T=\boldsymbol{O}$, the approximate equations become exact.
    \begin{proof}
        At time $t$, we parametrize group elements by $g(\boldsymbol{x})=\mu(t)\exp(\boldsymbol{x})$. We first convert the non-parametric SDE (\ref{eq:prop_nonpara_SDE}) to a parametric Ito's SDE using Theorem \ref{thm:Ito_equi}:
\begin{equation}
\boldsymbol{x}(t+\Delta t)=\boldsymbol{x}(t)+J_r^{-1}\tilde{\boldsymbol{h}}\Delta t+J_r^{-1}\tilde{H}\Delta \boldsymbol{W}.
\end{equation}
Denote the probability density function of $\boldsymbol{x}(t)$ and $\boldsymbol{x}(t+\Delta t)$ as $\tilde{p}_{\mu}(\boldsymbol{x},t)$ and $\tilde{p}_{\mu}(\boldsymbol{x},t+\Delta t)$. The mean and covariance of $\boldsymbol{x}(t+\Delta t)$ are
\begin{equation}
\begin{aligned}
\langle \boldsymbol{x}(t+\Delta t)\rangle&=\langle \boldsymbol{x}(t)+J_r^{-1}\tilde{\boldsymbol{h}}\Delta t+J_r^{-1}\tilde{H}\Delta \boldsymbol{W}\rangle\\
&=\langle J_r^{-1}\tilde{\boldsymbol{h}}\rangle \Delta t,
\end{aligned}
\end{equation}
\begin{equation}
\begin{aligned}
\Sigma'(t+\Delta t)&=\langle \boldsymbol{x}(t+\Delta t)\boldsymbol{x}^T(t+\Delta t)\rangle\\
&\qquad-\langle \boldsymbol{x}(t+\Delta t)\rangle\langle \boldsymbol{x}^T(t+\Delta t)\rangle\\
&= \Sigma(t)+\langle J_r^{-1}\tilde{H}\tilde{H}^TJ_r^{-T}\rangle \Delta t \\
&\qquad +\text{sym}(\langle J_r^{-1}\tilde{\boldsymbol{h}}\boldsymbol{x}^T\rangle)\Delta t+\mathcal{O}(\Delta t^2)
\end{aligned}
\end{equation}
where we have used $\Delta W_i \cdot \Delta W_j=\delta_{ij} \Delta t$ and that $\Delta \boldsymbol{W}= \boldsymbol{W}(t+\Delta t)-\boldsymbol{W}(t)$ and $\boldsymbol{x}(t)$ are independent. Now we have obtained the mean and covariance for $\tilde{p}_{\mu}(\boldsymbol{x},t+\Delta t)$ and proceed to calculate the group-theoretic mean $\boldsymbol{m}$ and covariance $\Sigma(t+\Delta t)$ of $g(\boldsymbol{x}(t+\Delta t))=\mu(t)\exp(\boldsymbol{x}(t+\Delta t))$. 

In applying Theorem \ref{thm:fit}, the support of $\tilde{p}_{\mu}(\boldsymbol{x},t+\Delta t)$ should be within $D$ and the expectation should be calculated using $\tilde{p}_{\mu}(\boldsymbol{x},t+\Delta t)$. 
The former is only true when $HH^T=\boldsymbol{O}$.
Nevertheless, when $HH^T$ and $\Delta t$ are very small, we assume the former approximately holds.
For the latter,
we can use the approximation that $\int v(\boldsymbol{x})\tilde{p}_{\mu}(\boldsymbol{x},t+\Delta t)d\boldsymbol{x}=\int v(\boldsymbol{x})\tilde{p}_{\mu}(\boldsymbol{x},t)dt+\mathcal{O}(\Delta t)$ to estimate the expectation by $\tilde{p}_{\mu}(\boldsymbol{x},t)$. 
The $\langle\cdot \rangle$ below denotes the expectation using $\tilde{p}_{\mu}(\boldsymbol{x},t)$.
Using Theorem \ref{thm:fit}, the group-theoretic mean of $g(\boldsymbol{x}(t+\Delta t))$ is estimated by
\begin{equation}
    \mu(t+\Delta t)=\mu(t)\exp(\boldsymbol{m}')
\end{equation}
where
\begin{equation}
\begin{aligned}
\boldsymbol{m}'&\,{\approx}\,\big(\langle J_l^{-1}\rangle^{-1}+\mathcal{O}(\Delta t)\big)\big(\langle J_r^{-1}\tilde{\boldsymbol{h}}\rangle\Delta t+\mathcal{O}(\Delta t^2)\big)\\
&\,{\approx}\,\langle J_l^{-1}\rangle^{-1}\langle J_r^{-1}\tilde{\boldsymbol{h}}\rangle\Delta t+\mathcal{O}(\Delta t^2),\\
\end{aligned}
\end{equation}
and the covariance matrix is estimated by
\begin{equation}
\begin{aligned}
\Sigma(t+\Delta t)&\,{\approx}\,\Sigma(t)+\langle J_r^{-1}\tilde{H}\tilde{H}^TJ_r^{-T}\rangle \Delta t \\
&\,\, +\text{sym}\big(\langle J_r^{-1}\tilde{\boldsymbol{h}}\boldsymbol{x}^T\!-\!J_l^{-1}\boldsymbol{m}'\boldsymbol{x}^T\rangle\big) \Delta t+\mathcal{O}(\Delta t^2).
\end{aligned}
\end{equation}
By substituting the expression for $\tilde{\boldsymbol{h}}$ and $\tilde{H}$ in Theorem \ref{thm:Ito_equi} and taking the limit $\Delta t \rightarrow 0$, we arrive at the propagation equations.
    \end{proof}
\end{theorem}

After having the mean and covariance, we can construct a concentrated Gaussian distribution $g=\mu(t)\exp(\boldsymbol{x})$, $\boldsymbol{x}\sim \mathcal{N}(\boldsymbol{0},\Sigma(t))$ as an approximate probability distribution.
When a new observation is obtained, it can be used as a prior distribution for Bayesian fusion.
We proceed to derive the fusion method.

\subsection{Bayesian Fusion}
Given a prior distribution on $G$, $g= \mu \exp(\boldsymbol{x}), \boldsymbol{x}\sim\mathcal{N}(\boldsymbol{0},P^{-})$, and an observation model $g_z=k(g)\exp(\boldsymbol{r}),\boldsymbol{r}\sim \mathcal{N}(\boldsymbol{0},R)$, where $k:G_1\rightarrow G_2$ and $G_2$ is a Lie group, we aim to calculate the group-theoretic mean and covariance of the posterior probability distribution $p(g|g_z)$.
When the group $G_2$ is $\mathbb{R}^M$, the observation model reduces to the common observation model on Euclidean space $\boldsymbol{g}_z=\boldsymbol{k}(g)+\boldsymbol{r}$.
Another case where $G_2=G_1$ and $k(g)=g$ is also of practical interest.
We first present a fusion method for the general problem and then state the simplification in these two cases.

\subsubsection{General Method}
Following the idea in Section \ref{sec:prop_eqs}, we use the parametrization $g(\boldsymbol{x})=\mu \exp(\boldsymbol{x})$ and $g_z(\boldsymbol{z})=k(\mu)\exp(\boldsymbol{z})$.
The parametrization gives
\begin{equation}\label{eq:obs_z}
\begin{aligned}
\boldsymbol{z}&=\tilde{\boldsymbol{k}}(\boldsymbol{x},\boldsymbol{r})=\log^{\vee}\big[k(\mu)^{-1}k\big(\mu\exp(\boldsymbol{x})\big)\exp(\boldsymbol{r})\big].
\end{aligned}
\end{equation}
We first estimate the mean of the posterior of $\boldsymbol{x}$ on the parameter space and then estimate the group-theoretic mean and covariance.
The posterior $p(\boldsymbol{x}|g_z)$ can be calculated by the update step in a Gaussian filter \cite{sarkka2023bayesian}:
\begin{theorem}\label{thm:gaussian_filter_posterior}
    The mean $\boldsymbol{m}$ and covariance matrix $\Sigma$ of the posterior distribution $p(\boldsymbol{x}|g_z)$ is
    \begin{equation}\label{eq:gaussian_filter_posterior}
        \begin{aligned}
            \tilde{\boldsymbol{m}}&=\langle\tilde{\boldsymbol{k}}(\boldsymbol{x},\boldsymbol{r})\rangle\\
            S&=\big\langle \big( \tilde{\boldsymbol{k}}(\boldsymbol{x},\boldsymbol{r})-\tilde{\boldsymbol{m}}\big)\big(\tilde{\boldsymbol{k}}(\boldsymbol{x},\boldsymbol{r})-\tilde{\boldsymbol{m}} \big)^T\big\rangle\\
            C&=\langle \boldsymbol{x}\big(\tilde{\boldsymbol{k}}(\boldsymbol{x},\boldsymbol{r})-\tilde{\boldsymbol{m}}\big)^T\rangle \\
            K&=CS^{-1}\\
            \boldsymbol{m}&=K\big(\log^{\vee}(k^{-1}(\mu)g_z)-\tilde{\boldsymbol{m}}\big)\\
            \Sigma &= P^{-}-KSK^T.
        \end{aligned}
    \end{equation}
    where $\langle v(\boldsymbol{x},\boldsymbol{r}) \rangle \doteq \int v(\boldsymbol{x},\boldsymbol{r})\mathcal{N}(\boldsymbol{x}|\boldsymbol{0},P^-)\mathcal{N}(\boldsymbol{r}|\boldsymbol{0},R)d\boldsymbol{x}d\boldsymbol{r}$.
\end{theorem}
After haveing $p(\boldsymbol{x}|g_z)$, we apply Theorem \ref{thm:fit}:
\begin{theorem}
    The group-theoretic mean $\mu_m$ and covariance matrix $\Sigma_m$ for the posterior $p(g|g_z)$ are
    \begin{equation}\label{eq:fusion_general}
        \begin{aligned}
        &\boldsymbol{m}'=\langle J_l^{-1} \rangle^{-1} \boldsymbol{m} \\
        &\mu_m \approx \mu\exp (\boldsymbol{m}'+\mathcal{O}(|\boldsymbol{m}'|^2))\\
        &\Sigma_{m}\approx \Sigma-\text{sym}\big(\langle J_l^{-1}\boldsymbol{m}'\boldsymbol{x}^T \rangle\big)
+\mathcal{O}(|\boldsymbol{m}'|^2)
        \end{aligned}
    \end{equation}
    where $\langle v(\boldsymbol{x})\rangle\doteq\int v(\boldsymbol{x})\mathcal{N}(\boldsymbol{x}|\boldsymbol{0},\Sigma)d\boldsymbol{x}$, $sym(A)\doteq A+A^T$, and $\boldsymbol{m}$ and $\Sigma$ are from equation (\ref{eq:gaussian_filter_posterior}).
    \begin{proof}
        The results come from the direct application of Theorem \ref{thm:fit}.
        There is one technical issue that makes equality become approximation: the non-linearity of $k(g)$ can lead to a highly non-Gaussian $p(\boldsymbol{x}|g_z)$, which makes the expectation calculation in Theorem \ref{thm:fit} intractable.
        Here, we approximate $p(\boldsymbol{x}|g_z)$ by constructing a Gaussian distribution using $\boldsymbol{m}$ and $\Sigma$.
        The reason why we use $\langle v(\boldsymbol{x})\rangle\doteq\int v(\boldsymbol{x})\mathcal{N}(\boldsymbol{x}|\boldsymbol{0},\Sigma)d\boldsymbol{x}$ instead of $\langle v(\boldsymbol{x})\rangle\doteq\int v(\boldsymbol{x})\mathcal{N}(\boldsymbol{x}|\boldsymbol{m},\Sigma)d\boldsymbol{x}$ is that this substitution causes $\mathcal{O}(|\boldsymbol{m}'|^2)$ level error when calculating $\langle J_l^{-1} \rangle^{-1}\boldsymbol{m}$ and $\langle J_l^{-1}\boldsymbol{m}'\boldsymbol{x}^T\rangle $, which is of the same order of error as in Theorem \ref{thm:fit}.
    \end{proof}
\end{theorem}

The minimizer property of the estimated mean is stated by the following theorem.
The proof uses the definition of the means and the idea of the proof for Theorem \ref{thm:mean_eq}.
\begin{theorem}\label{thm:post_min}
    The group-theoretic mean $\mu_G(g_z)$ of the posterior $p(g|g_z)$ is the minimizer of the cost functional
    \begin{equation}
        c_1(\mu) = \bigg|\int_{G_1,G_2} \log^{\vee}\big( \mu^{-1}(g_z)g \big)p(g_z,g)dg dg_z \bigg|_2^2
    \end{equation}
    and the Fr\'echet mean $\mu_F(g_z)$ is the minimizer of 
    \begin{equation}
        c_2(\mu) = \int_{G_1,G_2} \big|\log^{\vee}\big( \mu^{-1}(g_z)g \big)\big|^2_2 p(g_z,g)dg dg_z.
    \end{equation}
    When $G_1=SO(3)$, the minimizer of these two costs is the same.
\end{theorem}

\subsubsection{Case 1: $G_2=\mathbb{R}^M$}
In this case, the observation $\boldsymbol{g}_z\in \mathbb{R}^M$ is a vector.
The equation (\ref{eq:obs_z}) reduces to ${\boldsymbol{z}}=\tilde{\boldsymbol{k}}(\boldsymbol{x},\boldsymbol{r})=\boldsymbol{k}(\mu \exp(\boldsymbol{x}))-\boldsymbol{k}(\mu)+\boldsymbol{r}$.
When expanding terms in equation (\ref{eq:gaussian_filter_posterior}) and (\ref{eq:fusion_general}) using Taylor's expansion on Lie groups and keeping terms up to the second-order, we have the following fusion formula:
\begin{corollary}\label{coro:euc}
When $G_2=\mathbb{R}^M$, the group-theoretic mean $\mu_m$ and covariance matrix $\Sigma_m$ for the posterior $p(g|\boldsymbol{z})$ are estimated by
\begin{equation}
    \begin{aligned}
            S&=P_{ij}^{-}(E^r_i \boldsymbol{k})(E^r_j \boldsymbol{k})^T+R\\
            C&=P^{-}\boldsymbol{e}_i (E_i^r \boldsymbol{k})^T \\
            K&=CS^{-1}\\
            \boldsymbol{m}&=K\big(\boldsymbol{g}_z-\boldsymbol{k}(\mu)-\frac{1}{2}P_{ij}^{-}(E_i^rE_j^r \boldsymbol{k}) \big)\\
            \Sigma &= P^{-}-KSK^T\\
                    \boldsymbol{m}'&=\bigg(\mathbb{I}-\frac{1}{12}\Sigma_{ij}ad_i ad_j\bigg) \boldsymbol{m} \\
        \mu_m &= \mu\exp (\boldsymbol{m}')\\
        \Sigma_{m}&= \Sigma+\text{sym}\bigg(\frac{1}{2}\Sigma_{ij}ad_i \boldsymbol{m}'\boldsymbol{e}_j^T \bigg),
    \end{aligned}
\end{equation}
where $sym(A)\doteq A+A^T$, $ad_i\doteq [ad_{E_i}]$ and other Lie operators are for $G_1$.
\end{corollary}
\noindent We have neglected terms that contains $P_{ij}^{-}P_{mn}^{-}$ in the above formula different from the second-order EKF in \cite{sarkka2023bayesian}.

\subsubsection{Case 2: $G_1=G_2$}
In this case, the equation (\ref{eq:obs_z}) becomes $\boldsymbol{z}=\tilde{\boldsymbol{k}}(\boldsymbol{x},\boldsymbol{r})=\log^{\vee}\big(\exp(\boldsymbol{x})\exp(\boldsymbol{r})\big)$.
Using the Baker-Campbell-Hausdorff
\begin{equation}
\begin{aligned}
\log^{\vee}\!\big(\exp(\boldsymbol{x})\exp(\boldsymbol{r})\big)&=\boldsymbol{x}+\boldsymbol{r}+ad_{\boldsymbol{x}^{\wedge}}\boldsymbol{r}+\frac{1}{12}(ad_{\boldsymbol{x}^{\wedge}}ad_{\boldsymbol{x}^{\wedge}}\boldsymbol{r}
\\
& \qquad+ad_{\boldsymbol{r}^{\wedge}}ad_{\boldsymbol{r}^{\wedge}}\boldsymbol{x})+ \mathcal{O}(|\boldsymbol{x}|^2\cdot|\boldsymbol{r}|^2)\\ 
&\qquad\quad+\mathcal{O}(|\boldsymbol{x}|^3,|\boldsymbol{r}|^3),
\end{aligned}
\end{equation}
and neglecting higher order terms (denoted by $\mathcal{O}(\cdot)$), we have the fusion formula to be
\begin{corollary}\label{coro:group}
When $G_1=G_2$ and $k(g)=g$, the group-theoretic mean $\mu_m$ and covariance matrix $\Sigma_m$ for the posterior $p(g|\boldsymbol{z})$ are estimated by
\begin{equation}
    \begin{aligned}
            \boldsymbol{m}&=P^{-}(P^{-}\!+R)^{-1}\log^{\vee}(\mu^{-1}g_z)\\
            \Sigma &= P^{-}-P^{-}(P^{-}\!+R)^{-1}P^{-}\\
                    \boldsymbol{m}'&=\bigg(\mathbb{I}-\frac{1}{12}\Sigma_{ij}ad_i ad_j\bigg) \boldsymbol{m} \\
        \mu_m &= \mu\exp (\boldsymbol{m}')\\
        \Sigma_{m}&= \Sigma+\text{sym}\bigg(\frac{1}{2}\Sigma_{ij}ad_i \boldsymbol{m}'\boldsymbol{e}_j^T \bigg).
    \end{aligned}
\end{equation}
where $sym(A)\doteq A+A^T$.
\end{corollary}


\section{Numerical Experiments} \label{sec:Results}

Since the accuracy of the propagation equation has been demonstrated in \cite{ye2023uncertainty}, we only test the fusion method.
We consider two observation models for the 3D attitude estimation problem in which the state space is $SO(3)$ and the observation lives in Euclidean space and $SO(3)$.
The first one uses the gravitational and magnetic measurement \cite{brossard2017unscented} and the second one observes the full state directly.

\subsection{Observation Model Description}


\subsubsection{3D attitude estimation with gravitational and magnetic measurement}\label{sec:exp_3d_euc}
Assume a prior distribution of the ground-truth state is given by $R=\exp(\boldsymbol{\xi})\exp(\boldsymbol{v})$, where $\boldsymbol{\xi}=(\frac{1}{3}\pi,\frac{1}{4}\pi,\frac{1}{6}\pi)^T$ and  $\boldsymbol{v}\sim \mathcal{N}(0,\text{diag}(0.5,1,0.8))$. The measurement model is $\boldsymbol{g}_z(R)=(\boldsymbol{g}^T R,\boldsymbol{b}^T R)^T+\boldsymbol{r}$, where $\boldsymbol{g}=(0,0,-9.82)^T$, $\boldsymbol{b}=(0.33,0,-0.95)^T$ and $\boldsymbol{r}\sim \mathcal{N}(\boldsymbol{0},Q)$.
The covariance matrix takes the form $Q=\tau\cdot\text{diag}(0.3,0.3,0.3,0.1,0.1,0.1)$, where we test different sets of the coefficient $\tau$ from $1\times 10^{-3}$ to $1$.

\subsubsection{3D attitude estimation with state measurement}
Assume the prior distribution is the same as the one in Section \ref{sec:exp_3d_euc} above. 
The observation model is $g_z=R\exp(\boldsymbol{r})\in SO(3)$, where $\boldsymbol{r}\sim \mathcal{N}(\boldsymbol{0}, Q)$. 
The covariance matrix takes the form $Q=\tau\cdot\text{diag}(0.3,0.3,0.3)$, where we test with $\tau$ from $1\times 10^{-3}$ to $1$.

\subsection{Sampling and Evaluation Metric}
We sample from the prior distribution to form the set of ground-truth state $R_i$ and use the observation model to sample observations $[{g_z}]_i$ for each $R_i$.
In total, we draw $N=10,000$ samples, $\{(R_i,{[{g}_z}]_i)\,\,|\,\,i=1,2,...,N\}$.
We evaluate the goodness of the estimated mean $\mu(g_{z})$ by two cost functions from Theorem \ref{thm:post_min}:
\begin{equation}
\begin{aligned}
c_1(\mu)
&\approx \bigg|\frac{1}{N}\sum_{i=1}^N \log^{\vee}(R^{-1}_i \mu([{g_z}]_i)) \bigg|_2^2\\
c_2(\mu)
    &\approx\frac{1}{N}\sum_{i=1}^N |\log^{\vee}\big(R_i^{-1}\mu([{g_z}]_i)\big)|_2^2.
\end{aligned}
\end{equation}
Our method is compared against the update step of a Kalman filter, \textit{i}.\textit{e}. the formula in Corollary \ref{coro:euc} and Corollary \ref{coro:group} without the modification step.

\subsection{Experiment Results}
\begin{figure}[thpb] 
      \centering
      \includegraphics[width=0.35\textwidth]{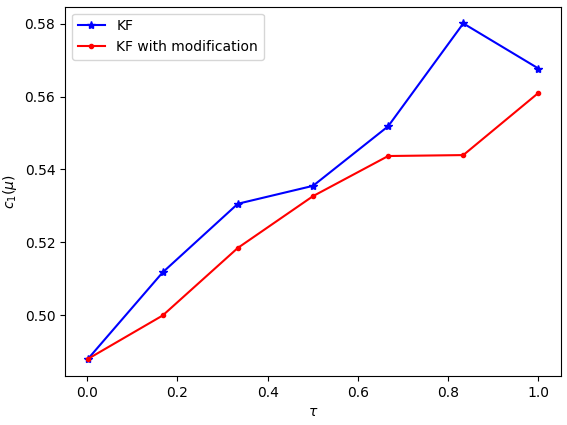}
      \caption{The plot of $c_1(\mu)$ for observation model 1 (on Euclidean space).}
      \label{fig:1}
\end{figure}

\begin{figure}[thpb]
      \centering
      \includegraphics[width=0.35\textwidth]{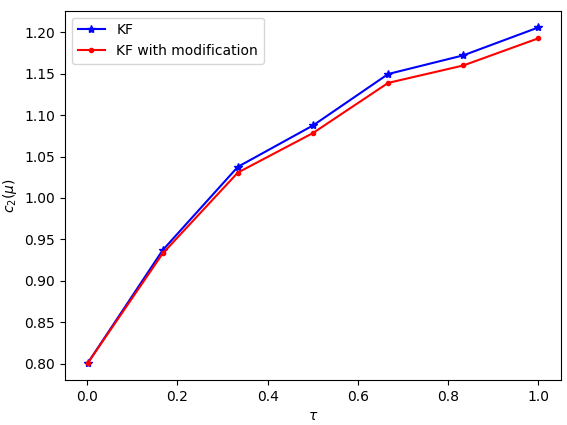}
      \caption{The plot of $c_2(\mu)$ for observation model 1 (on Euclidean space).}
      \label{fig:2}
\end{figure}

\begin{figure}[thpb]
      \centering
      \includegraphics[width=0.35\textwidth]{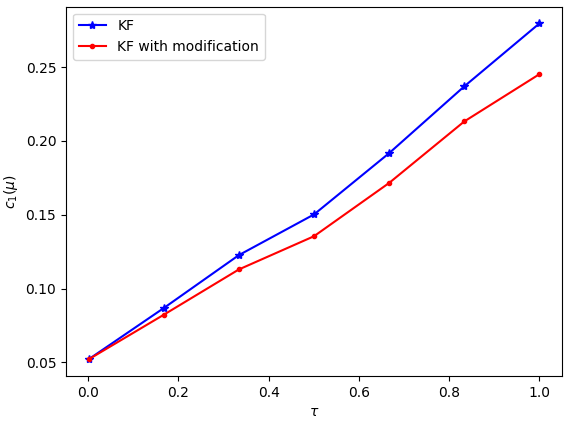}
      \caption{The plot of $c_1(\mu)$ for observation model 2 (on Lie group).}
      \label{fig:3}
\end{figure}
\begin{figure}[thpb]
      \centering
      \includegraphics[width=0.35\textwidth]{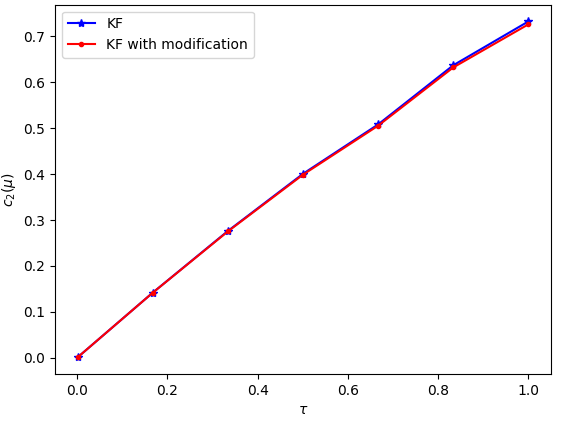}
      \caption{The plot of $c_2(\mu)$ for observation model 1 (on Lie group).}
      \label{fig:4}
\end{figure}

The experiment results shown in Figure \ref{fig:1}-\ref{fig:4} indicate that the modification term improves the fusion estimate evaluated by two metrics and for two observation models.


\section{Conclusions \& Future Work}\label{sec:Concl}

This paper provides a derivation to the mean and covariance propagation equations when the dynamics model is a stochastic differential equation (SDE) on Lie groups and a simple modification to the update step of a Kalman filter.
We first derive the relationship between a non-parametric SDE defined by Mckean-Gangolli injection and a parametric SDE on the exponential coordinate of the group.
Then we derive a mean and covariance fitting formula for probability distributions on Lie groups that are defined by projecting a concentrated distribution on the exponential coordinate to the group.
Combining these two tools, we derive the mean and covariance propagation equations for a non-parametric SDE.
We proceed to use the fitting formula to derive a modification term to the update step of a Gaussian/Kalman filter.
Experiment results show the modification term improves the posterior mean estimate.

In the future, we will analyze the convergence property of a filter based on our propagation equation and fusion method.
We may also extend to propagating a multimodal probability distribution on Lie groups which has been considered in Euclidean space \cite{lambert2022variational}.





\section{Appendix}\label{sec:app}
\subsection{A useful Taylor's expansion formula}
\begin{equation}
    \begin{aligned}
&\qquad\log^{\vee}(\exp(\boldsymbol{-\epsilon}^{\wedge})\exp(\boldsymbol{x}^{\wedge}))\\
&=\boldsymbol{x}+\epsilon_iE^l_i\boldsymbol{x}+\frac{1}{2}\epsilon_i \epsilon_j E_i^l E_j^l \boldsymbol{x}+\mathcal{O}(\boldsymbol{\epsilon}^3)\\
&=\boldsymbol{x}-J_l^{-1}\boldsymbol{\epsilon}+\frac{1}{2}\frac{\partial J_l^{-1}}{\partial x_k}\boldsymbol{\epsilon}\boldsymbol{\epsilon}^TJ_l^{-T}\boldsymbol{e}_k+\mathcal{O}(\boldsymbol{\epsilon}^3)
    \end{aligned}
\end{equation}


\subsection{Proof of Lemma \ref{lemma:Ito_Stra_equi}}
\begin{proof}
    The proof goes the same way as in \cite{gardiner1985handbook} for SDEs on Euclidean space.
    Suppose the solution to the Stratonovich's SDE is equivalent to the Ito's SDE. 
    We approximate terms evaluated at $t=t+\frac{1}{2}dt$ using Taylor's expansion on Lie groups and substitute the Ito's SDE into the Stratonovich's SDE to derive the condition:
    \begin{equation}\small
        \begin{aligned}
g(t+dt)&=g(t)\exp\big(\boldsymbol{h}^s\Bigr|_{\tiny\substack{g=g(t)\\t=t}}dt+H^s\Bigr|_{\tiny\substack{g=g(t+\frac{1}{2}dt)\\t=t+\frac{1}{2}dt}}d\boldsymbol{W}\big)\\
&=g(t)\exp\bigg[\boldsymbol{h}^s\Bigr|_{\tiny\substack{g=g(t)\\t=t}}dt+H^s\Bigr|_{\tiny\substack{g=g(t)\\t=t}}d\boldsymbol{W}\\&\qquad+\sum_{i,\ j} E^r_i({H}^s){H}_{ij}\Bigr|_{\tiny\substack{g=g(t)\\t=t}}\big({W}_j(t+\frac{dt}{2})-{W}_j(t)\big)\\
&\qquad\cdot (\boldsymbol{W}(t+dt)-\boldsymbol{W}(t))+\mathcal{O}(dt^{3/2}) \bigg]\\
&=g(t)\exp\bigg[\boldsymbol{h}^s\Bigr|_{\tiny\substack{g=g(t)\\t=t}}dt+H^s\Bigr|_{\tiny\substack{g=g(t)\\t=t}}d\boldsymbol{W}\\
&\qquad+\frac{dt}{2} E^r_i({H}^s_{kj}){H}_{ij}\Bigr|_{\tiny\substack{g=g(t)\\t=t}}\boldsymbol{e}_k+\mathcal{O}(dt^{3/2}) \bigg]\\
\end{aligned}
    \end{equation}
    where we have used the fact that $(W_i(t+dt)-W_i(t+\frac{1}{2}dt))(W_j(t+\frac{1}{2}dt)-W_j(t))=0$  and that $(W_i(t+dt)-W_i(t+\frac{1}{2}dt))(W_j(t+dt)-W_j(t+\frac{1}{2}dt))=\delta_{ij}\cdot \frac{dt}{2}$ in the sense of mean squared limit. 
    Since this equation is equivalent to the Ito's SDE, we arrive at the equivalence condition after comparing terms. 
\end{proof}

\bibliography{References}

\end{document}